\documentstyle[aps,preprint]{revtex}
\begin{document}
\draft
\preprint{MA/UC3M/07/94}

\title{Dynamics of electric field domains and oscillations of
the photocurrent in a simple superlattice model}

\author{L. L. Bonilla, J. Gal\'{a}n\cite{jorge}, J. A. Cuesta, F. C.
Mart\'{\i}nez  and J. M. Molera}
\address{
Escuela Polit\'{e}cnica Superior\\
Universidad Carlos III de Madrid\\
Butarque 15; 28911 Legan\'{e}s, Spain}
\date{\today}

\maketitle
\begin{abstract}
A discrete model is introduced to account for the time-periodic
oscillations of the photocurrent in a superlattice observed by Kwok et
al, in an undoped 40 period AlAs/GaAs superlattice. Basic ingredients
are an effective negative differential resistance due to the sequential
resonant tunneling of the photoexcited carriers through the potential
barriers, and a rate equation for the holes that incorporates
photogeneration and recombination. The photoexciting laser acts as a
damping factor ending the oscillations when its power is large enough.
The model explains: (i) the known oscillatory static I-V characteristic
curve through the formation of a domain wall connecting high and low
electric field domains, and (ii) the photocurrent and photoluminescence
time-dependent oscillations after the domain wall is formed. In our
model, they arise from the combined motion of the wall and the shift
of the values of the electric field at the domains. Up to a certain
value of the photoexcitation, the non-uniform field profile with two
domains turns out to be metastable: after the photocurrent
oscillations have ceased, the field profile slowly relaxes toward
the uniform stationary solution (which is reached on a much longer
time scale). Multiple stability of stationary states and hysteresis are
also found. An interpretation of the oscillations in the
photoluminescence spectrum is also given.

\bigskip

\noindent Key words: current instabilities; electric field domains;
metastability; dynamical effects in superlattices.

\end{abstract}
\pacs{Pacs numbers: 72.20.Ht., 05.45+b, 85.30.Fg}

\setcounter{equation}{0}

\narrowtext

\section{Introduction}
\label{sec-introduction}

In recent experiments by Kwok et al. the time-dependent transport
properties of photoexcited undoped superlattices were investigated
\cite{merlin}. In a typical example, an undoped 40 period AlAs/GaAs
superlattice was mounted in a {\em p-i-n} diode and continuously
illuminated by laser light at 4K. When the laser power was in a certain
interval and the applied {\em dc} voltage bias was large enough, damped
time-dependent oscillations of the photocurrent (PC) and the peaks in
the photoluminescence (PL) spectrum were observed. The applied fields
in the experiment \cite{merlin} are high, the potential barriers are
wide (between 30 and 40 \AA), and the minibands correspondingly narrow
so that the coherence length is comparable or smaller than the width of
one quantum well. Then the quantum wells in the superlattice are weakly
coupled and formation of electric field domains may appear
\cite{cavicchi,grahn89,grahn90,kwok91,grahn92}. The subject of
formation and expansion of electric field domains in (doped)
superlattices is an old one and it goes back to the conductance
measurements of Esaki and Chang in 1974. \cite{esaki} Typically the
dominant transport process in each domain is resonant tunneling between
different subbands of the wells belonging to the domain.

Time-dependent oscillations of the PC in undoped illuminated
superlattices have also been observed by Le-Person et al.
\cite{leperson} for quite different superlattices where miniband
transport is prevalent (narrow barriers, so that the quantum wells are
not weakly coupled. In these conditions it is not clear whether
formation and propagation of electric field domains are relevant).
These measurements were interpreted in terms of Gunn dipole domains of
a classical drift-diffusion model.

In experiments with weakly coupled superlattices at high electric
fields, the dominant transport mechanism is resonant tunneling: It
follows from Kazarinov and Suris's tight binding calculation
\cite{kazar} that when the Wannier-Stark level splitting is smaller
than the distance between the two lowest subbands of each quantum well
of the superlattice, the electron is localized and the current through
the structure is small. When both quantities are comparable, there is
a peak in the current-field curve due to resonant tunneling between the
first subband, $e_1$, of one well and the second subband, $e_2$, of the
next one (let us denote by $e_1\rightarrow e_2$ this kind of resonant
tunneling). Further peaks correspond to other resonant transitions
between different energy subbands of the wells. As a result of resonant
tunneling, electrons populate higher subbands of the wells and
PL measurements may be used to probe the change
of electron population in higher subbands. \cite{grahn89,grahn90} In
superlattices with wide wells the resonant tunneling process is
sequential, which means that after a tunneling event the electron decay
very fast to a lower (e.g., the first) subband due to scattering.
\cite{capasso} For the superlattices described above, the intersubband
scattering time is of the order of 1 ps which is very small compared to
the typical tunneling time of 100 ps. \cite{grahnprl90} When electric
field domains are formed, the wells belonging to each domain undergo
similar sequential resonant tunneling (SRT) events. In Ref.
\onlinecite{merlin} and previously in Ref. \onlinecite{grahn90}, three
domains are observed for different bias and illumination: domain I
where SRT is $e_1\rightarrow e_1$ (the subbands of all the wells in
this domain are aligned at zero electric field), domain II with
$e_1\rightarrow e_2$ SRT, and domain III with $e_2\rightarrow e_3$ or
$e_1\rightarrow e_3$ SRT according to the sample. We have depicted
schematically the SRT processes at a sample where domains II and III
coexist in Fig.\ \ref{jumps}. Time-dependent PC oscillations are
observed as a transient toward a stationary state where domains II and
III coexist.\cite{merlin}

\bigskip

In this paper we describe and analyze a simple discrete drift model
that explains several features of the experiments reported in Ref.
\onlinecite{merlin}, particularly the stationary current-voltage
characteristic curve \cite{grahn90} and the time-dependent oscillations
of the PC in the region of the parameter space where domains
II and III coexist. Previous theoretical works on electric field
domains in superlattices include Grahn et al's \cite{grahn90}
characterization of stationary states, Laikhtman's equivalent circuit
model \cite{laik} and Korotkov et al's simulations based upon the
analogy of a slim superlattice with a one-dimensional array of metallic
tunnel junctions. \cite{korotkov} We shall comment on these works in
the Discussion. To motivate our model let us consider several important
time scales for the superlattices object of our study. First of all,
the characteristic time scale of the PC oscillations in Ref.
\onlinecite{merlin} is of 100 ns. The time scale for carrier
thermalization is of 0.1 ps while the carriers reach thermal
equilibrium with the lattice after a time that ranges from 1 to 100 ps.
\cite{shah} This means that
in time scales of the order of nanoseconds (the experimental time
scale), we may consider the holes and electrons to be at local
equilibrium within each quantum well $j$ at the lattice temperature,
and with given values of their densities, $\tilde{p}_{j}$ and
$\tilde{n}_{j}$. The process of reaching a stationary state might be
seen as the attempt of reaching a ``global equilibrium'' starting from
``local equilibrium'' through tunneling processes that communicate
different quantum wells, self-consistency of the electric field and
scattering and interband processes. In this spirit we consider the
quantum wells as entities characterized by average values of the
electric field, $\tilde{E}_{j}$ for the $j$-th well, and of the
densities of holes and electrons, $\tilde{p}_{j}$ and $\tilde{n}_{j}$,
respectively. We then propose the following transport equations to
describe the dynamics of the superlattice:
\begin{eqnarray}
\tilde{E}_{j} - \tilde{E}_{j-1} = \frac{e\,\tilde{l}}{\epsilon}\,
(\tilde{n}_{j} - \tilde{p}_{j}),                   \label{poisson}\\
\epsilon\,\frac{d \tilde{E}_{j}}{d \tilde{t}} +
e\,\tilde{v}(\tilde{E}_{j})\,\tilde{n}_{j} = \tilde{J},
                                                   \label{ampere} \\
\frac{d \tilde{p}_{j}}{d \tilde{t}} = \tilde{\gamma}
- \tilde{r}\,\tilde{n}_{j}\,\tilde{p}_{j},         \label{rate}
\end{eqnarray}
where $j= 1,\dots, N$. In this model:

\begin{enumerate}

\item Equation (\ref{poisson}) is the discrete Poisson equation
relating the field at two adjacent wells and the electric charge; $e$
is the electron charge, $\tilde{l}$ is the length of one period of the
superlattice, and $\epsilon$ is an average permittivity. Equation
(\ref{poisson}) may be considered as the average of the usual Poisson
equation over one period of  the superlattice. Since the
heterostructure corresponds to an intrinsic semiconductor, we do not
include charged impurities in (\ref{poisson}). Poisson's equation for
the $j=1$ well contains the field at the ``zeroth'' well, which
corresponds to the doped semiconductor {\em before}
the superlattice and it will be dealt with below.

\item Equation (\ref{ampere}) is Amp\`{e}re's law at the $j$-th well
establishing that the total current density $\tilde{J}$ is the sum of
the displacement current and the electron flux. We consider that the
holes are heavy and their contribution to the current is negligible.
\cite{capasso85} The electron flux $J_{j\rightarrow j+1}^{T}
= e\,\tilde{v}(\tilde{E}_{j})\,\tilde{n}_{j}$ is due to sequential
resonant tunneling between subbands of neighboring quantum wells as
said above. By writing this expression we are assuming that the
probability of tunneling from the well $j$ to the well $j+1$ is
proportional to the number of electrons in the well $j$, and we are
ignoring the small ``reverse tunneling'' from the well $j+1$ to the
well $j$. The probability of tunneling is larger for fields
corresponding to SRT (for which the subband $e_1$ of $j$ is aligned
with $e_2$ of $j+1$, or $e_2$ of $j$ with $e_3$ of $j+1$, etc).
This implies that the function $\tilde{v}(\tilde{E})$ (with the
dimensions of a velocity) has peaks at the corresponding values of the
electric field. A crucial assumption of our model is that the electron
``velocity'' corresponds to the static current--voltage characteristic
of the superlattice at low laser power (small $\tilde{\gamma}$; see
(\ref{const}) below). We take this curve as a datum of our model and
argue that our results do not depend qualitatively of the exact shape
of $\tilde{v}(\tilde{E})$. Since we are interested in the analysis of
the region where domains II ($e_1\rightarrow e_2$ SRT) and domains III
($e_2\rightarrow e_3$ or $e_1\rightarrow e_3$ SRT) coexist, we have
used a velocity curve with two local maxima centered in the same field
values as seen in the experiments. \cite{merlin,grahn90} The result is
Fig.\ \ref{ve}. To explore the region where domains I ($e_1\rightarrow
e_1$ SRT) and II coexist, it is convenient to add another (smaller)
peak to the velocity curve at zero electric field. The modifications
will be reported elsewhere.

\item Equation (\ref{rate}) is the rate equation for the hole
concentration. We consider only processes of photogeneration of
electron-hole pairs and recombination. The photogeneration term,
$\tilde{\gamma}$, is proportional to the power of the laser
(Ref. \onlinecite{seeger}, chapter 12; see also Delalande's paper in
Ref. \onlinecite{houches}), which we assume (for simplicity) that it
illuminates uniformly the superlattice. The recombination coefficient
is assumed to be constant.

\end{enumerate}

Notice that the rate equation for the electrons containing the
photogeneration, recombination and SRT processes is:
\begin{equation}
\frac{d \tilde{n}_{j}}{d \tilde{t}} = \tilde{\gamma}
- \tilde{r}\,\tilde{n}_{j}\,\tilde{p}_{j} + \frac{1}{\tilde{l}}
\{ \tilde{v}(\tilde{E}_{j-1})\,\tilde{n}_{j-1} -
\tilde{v}(\tilde{E}_{j})\, \tilde{n}_{j} \}.
\end{equation}
Here the SRT current coming from well $j-1$ to well $j$ brings
electrons to the $j$-th well at a rate
$\tilde{v}(\tilde{E}_{j-1})\,\tilde{n}_{j-1}/\tilde{l}$, while the SRT
current going from well $j$ to well $j+1$ detracts electrons from the
$j$-th well. This equation is equivalent to (\ref{ampere}), as it can
be shown by differentiating (\ref{poisson}) and using (\ref{ampere})
and (\ref{rate}).

In the $3N$ Equations (\ref{poisson}-\ref{rate}) there are $3N+2$
unknowns, $\tilde{J},\, \tilde{E}_{0}, \tilde{E}_{j}, \tilde{n}_{j},
\tilde{p}_{j}$, with $j= 1,\dots, N$. One additional equation is the
bias condition:
\begin{equation}
\frac{1}{N}\,\sum_{j=1}^{N} \tilde{E}_{j} = \frac{\tilde{\Phi}}{N\,
\tilde{l}}.
\label{bias}
\end{equation}
Here $\tilde{\Phi}$ is the difference between the applied voltage
and the built-in potential due to the doped regions outside the
superlattice (1.5 Volts in Ref. \onlinecite{merlin}).
$\frac{\tilde{\Phi}}{N\,\tilde{l}}$ is the average applied electric
field on the superlattice, which we will henceforth call the {\em
bias}. The missing condition is a boundary condition for the field at
the zeroth well,
$\tilde{E}_{0}$, ({\em before} the superlattice). We do not have direct
experimental evidence for what $\tilde{E}_{0}$ should be. Thus our
choice for $\tilde{E}_{0}$ has to be validated {\em a posteriori} by
comparing the results of our analysis with experiments and with the
consequences of a different choice. We shall use throughout this paper
the following boundary condition:
\begin{equation}
\tilde{E}_{0}(\tilde{t}) = \tilde{E}_{1}(\tilde{t}). \label{bc}
\end{equation}
This condition does not allow a charge build-up at the first
well and it is compatible with the experimental observation that the
field is the same for all the wells at low laser power:
\begin{equation}
\tilde{v}\left(\frac{\tilde{\Phi}}{N\tilde{l}}\right) =
\frac{\tilde{J}\,\tilde{r}^{1/2}}{e\,\tilde{\gamma}^{1/2}},\;\;
\tilde{E}_{j} = \frac{\tilde{\Phi}}{N\tilde{l}},\label{const}
\end{equation}
for all $j$. This equation says that the effective electron
velocity is (except for constant scale factors) the same as the static
I-V characteristic curve of the superlattice at low laser power, which
allows us inferring the form of $\tilde{v}(\tilde{E})$
from experimental data. Thus our boundary condition (\ref{bc}) forces
a uniform field profile (the same $\tilde{E}_j$ for all $j$) to be
{\em a solution of our model for any laser power}. As we shall see
later, this uniform solution is stable for low laser power whereas the
model {\em subject to the same boundary condition} (\ref{bc}) may have
stable non-uniform field profiles (with domains) for large enough laser
power. Further discussion of the boundary condition is postponed to
Section VI. For the velocity curve $\tilde{v}(\tilde{E})$, we
have used a variety of phenomenological curves with two
peaks and a negative differential resistance (NDR) region of negative
slope between them. We could have also obtained the velocity curve from
theories of static I-V curves for double barrier structures, but given
the qualitative insensitivity of our results to variations of the
velocity curve (as we will explain later), we have stuck to curves with
easy analytical expressions for the calculations presented here.

The boundary condition (\ref{bc}) allows for electric field domains
$E_j$, $j=1,\dots,N$ made of two constant values of the electric field
and a domain wall connecting them, as we shall discuss below. The fact
that our superlattice is embedded in a {\em p-i-n} diode implies that
domains with low field at $x=0$ and high field at $x=L$ are preferred
to domains with the symmetric configuration (see Eq. (\ref{poisson})).
Thus we shall only consider initial conditions that favor this kind of
electric field domain and we will not mention the domains that start at
$x=0$ with high electric field, even though they may be possible with
our boundary condition. When the superlattice forms part of a different
diode, e.g. {\em n-i-n}, the two kind of domains are allowed and richer
structures may arise according to the initial condition used in the
simulations.

\bigskip

It is convenient to render dimensionless the equations of our model
before we proceed to their analysis. For this, we adopt as the unit of
electric field that at the first maximum of the velocity curve,
$\tilde{E}_{M}\simeq 10^{5}$ V/cm. Then (\ref{poisson}) yields the
typical unit of electron density
\begin{equation}
\tilde{n} = \frac{\epsilon\tilde{E}_{M}}{e\tilde{l}}\simeq 10^{17}\,
\mbox{cm}^{-3},
\end{equation}
for $\tilde{l}\simeq 100$ \AA. There are two important time scales in
our model. From (\ref{ampere}) we see that the electrons employ a time
$\tau_{e} = \tilde{l}/\tilde{v}(\tilde{E}_{M})\simeq 0.4$ ns to tunnel
from one well to the next one of the superlattice. From (\ref{rate}),
we find the time scale in which the hole density varies
\begin{equation}
\frac{1}{\tau_{p}} = \tilde{r}\, \tilde{n}.
\end{equation}
Thus $\tau_{p}$ is essentially given by the recombination
coefficient $\tilde{r}$. $\tilde{r}$ is a decreasing function of the
electric field because the overlap between the electron and hole
wave functions decreases as the field increases. \cite{polland}
This accounts for a variation of $\tau_{p}$ from subnanoseconds to
tens of ns. \cite{polland} Here we adopt $\tau_{p} = 10$ ns which is
reasonable in view of the experimental values reported in Ref.
\onlinecite{merlin}. We measure the time in units  of $\tau_{p}$
which together with the previously mentioned units complete the
following definition of dimensionless variables:
\begin{eqnarray}
E_{j} = \frac{\tilde{E}_{j}}{\tilde{E}_{M}},\;\;
p_{j} = \frac{\tilde{p}_{j}}{\tilde{n}},\;\;
n_{j} = \frac{\tilde{n}_{j}}{\tilde{n}},\nonumber\\
t = \frac{\tilde{t}}{\tau_{p}} = \tilde{r}\,\tilde{n}\,\tilde{t},\;\;
J = \frac{\tilde{J}}{e\,\tilde{n}\,\tilde{v}(\tilde{E}_{M})},
\label{var}
\end{eqnarray}
and parameters
\begin{equation}
\phi = \frac{\tilde{\Phi}}{\tilde{E}_{M}\, N\tilde{l}},\;\;
\gamma = \frac{\tilde{\gamma}}{\tilde{r}\tilde{n}^{2}},\;\;
\beta = \frac{\tau_{e}}{\tau_{p}}
= \frac{\tilde{r}\tilde{l}\tilde{n}}{\tilde{v}(\tilde{E}_{M})}.
\label{par}
\end{equation}
Inserting (\ref{var},\ref{par}) into (\ref{poisson}--\ref{bc}) we
find the dimensionless system
\begin{eqnarray}
E_{j} - E_{j-1} &=& n_{j} - p_{j},                    \label{1}  \\
\beta\,\frac{d E_{j}}{d t} + v(E_{j})\, n_{j} &=& J,  \label{2}  \\
\frac{d p_{j}}{d t} &=& \gamma - n_{j}\, p_{j},       \label{3}  \\
\frac{1}{N}\,\sum_{j=1}^{N} E_{j} &=& \phi,           \label{4}  \\
E_{0} &=& E_{1}.                                      \label{5}
\end{eqnarray}
Here $\beta$ goes from $0.01$ to 1 and $\phi$ and $\gamma$ are the
dimensionless control parameters. These equations are to be solved
with initial conditions for the fields, $E_{j}(0)$, and the hole
concentrations, $p_{j}(0)$, compatible with the bias (\ref{4}) and the
boundary condition (\ref{5}). The initial conditions for the electron
density, $n_{j}(0)$, then follow from (\ref{1}).

\section{Steady states}
\label{sec-steady}

For given values of $\gamma$ and $J$ there are stationary solutions of
the equations (\ref{1}--\ref{3}) compatible with the boundary condition
(\ref{5}). Using equation (\ref{4}), we find the corresponding value
of $\phi$. From this we can obtain the curves $J=J(\phi)$
appearing in Fig.\ \ref{IVjm}. We have studied two different types of
stationary solutions, namely uniform solutions and two-domain
solutions.

\subsection{Uniform solutions}

By inserting $E_j\equiv E$ (constant) into (\ref{1}--\ref{4}), we find
\begin{eqnarray}
E & = & \phi                     \label{12}      \\
n_{j} & = & p_{j} = \sqrt{\gamma} \hspace{1.2cm} (j=1,\dots,N)
\label{11}  \\
J & = & \sqrt{\gamma}\,v(\phi)   \label{jhomo}
\end{eqnarray}
Equation (\ref{jhomo}) yield a static current-voltage characteristic
curve which is the same as $v(E)$ except for a scale factor. See Fig.\
\ref{IVjm}.

\subsection{Non-uniform two-domain solutions}

The stationary versions of equations (\ref{2}) and (\ref{3}) can be
solved for $n_j$ and $p_j$ in terms of the electric field $E_j$ and
then inserted in (\ref{1}). The result is the following discrete
mapping
\begin{equation}
E_{j-1} = f(E_j;\gamma,J)  \label{13}
\end{equation}
with
\begin{equation}
f(E;\gamma,J)\;\equiv E - \frac{J}{v(E)} + \frac{\gamma}{J}\, v(E).
\label{15}
\end{equation}
The mapping (\ref{13}) allows us to get all the electric field
profiles $\{ E_j,\; j = 1, \ldots, N \}$ that satisfy the boundary
condition $E_0\;=\;E_1$. From
these profiles and the bias condition (\ref{4}), we obtain the
current--voltage characteristic curve of Fig.\ \ref{IVjm}. Motivated
by experiments, we
are interested in stationary solutions with two electric field
domains, i.e. $E_j = E_L$\, ($E_L$ is the constant low field value)
for $j=1,\ldots,k $; $E_j = E_H$\, ($E_H$ is the constant high field
value) for $j=N-\tilde{k}+1,\ldots,N$, whereas for the remaining
quantum wells ($j=k+1,\dots,\tilde{k}$) $E_j$ is an increasing
function of $j$ that takes values on the interval $(E_L ,E_H )$. These
wells constitute the domain wall that separates domains II and III.

For these solutions to exist, a few requirements have to be fulfilled.
First of all, the boundary condition (\ref{5}) together with (\ref{13})
imply that $E_0$ should be a fixed point of the discrete mapping
(\ref{13}), i.e., a solution of
\begin{equation}
f(E;\gamma,J)=E  \label{14}
\end{equation}
(see Fig.\ \ref{f2}). Furthermore in order to have a non-uniform solution
 with domains, (\ref{14}) should have more than one fixed point. After
simplification, Eq. (\ref{14}) becomes (\ref{jhomo}) and therefore this
requirement is satisfied for certain values of $J$ described below. Let
$E_L$ be the fixed point of the mapping that we choose for $E_0$. With
the $v(E)$ profile depicted in Fig.\ \ref{ve}, (\ref{14}) may have one
or three solutions depending on the values of $J$ and $\gamma$. Let
$E_M$, $E_{M'}$ and $E_m$ be the positive electric field at the first
and second positive maximum and at the first positive minimum of
$v(E)$, respectively (see Fig.\ \ref{ve}). For
$0<J<\sqrt{\gamma}v(E_m)$ and for
$\sqrt{\gamma}v(E_M)<J<\sqrt{\gamma}v(E_{M'})$,
(\ref{14}) has only one solution on the interval $(0,E_{M'})$. For
$\sqrt{\gamma}v(E_m)<J<\sqrt{\gamma}v(E_M)$, there are three solutions
of (\ref{14}) on the same interval. Thus we can find a fixed point $E_L$
on the interval $(0,E_{M'})$ for $0<J<\sqrt{\gamma}v(E_{M'})$.
Secondly, a non-uniform profile should have $k < N$ wells with fields
equal to $E_L$ and $E_{k+1} > E_L$. This means that the equation
\begin{equation}
f(E;\gamma,J) =  E_L ,   \label{1iter}
\end{equation}
should have at least a solution different from the fixed point,
otherwise only the uniform solution described above is possible. This
second condition holds only for $\gamma > \gamma_1 \simeq 5.2 \times
10^{-3}$ (see Section \ref{sec-phase}). If one of the two lowest
fixed points of (\ref{13}) satisfies these two conditions, we can
select it as the value of the low field domain $E_L$ and let the $k$
first wells to have this value for their electric fields ($E_j=E_L$,
$\forall j=1,\dots,k$). Then one may choose as $E_{k+1}$ another
solution of (\ref{1iter}) different from $E_L$. The {\em p-i-n}
character of the diode prescribes that this is an acceptable solution
provided the electric field $E_{k+1}$ is higher than $E_L$. Then
(\ref{13}) implies that $E_j$ increases with $j$ and approaches another
fixed point, which is the value $E_H$ of the high field domain. As Fig.\
\ref{f2} shows, for each fixed $\gamma > \gamma_1$ both conditions are
satisfied only for values of $J$ on a certain interval and for some of
the fixed points of the mapping. The current-- applied bias ($J-\phi$)
characteristic curves obtained with the previously described method are
depicted in Fig.\ \ref{IVjm} for $\gamma = 0.016$. As explained before,
the continuous curve (together with the short-dashed part in the NDR
region) is $\sqrt{\gamma}\,v(\phi)$, with $v(E)$ depicted in Fig.\
\ref{ve}. It corresponds to the uniform solution,
obtained when all the $E_j$ stay on one of the fixed points (\ref{14}).
The $J$-$\phi$ curves for the non-uniform stationary solutions
with two domains explained above appear for $\phi$ on an
interval that ranges from a value slightly below $E_M$
to a value slightly below $E_{M'}$
in Fig.\ \ref{IVjm}. They exist for $\gamma > \gamma_1$ and $J$ on a
subinterval of $(\sqrt{\gamma}v(E_m),\sqrt{\gamma}v(E_{M}))$ whose
width increases with $\gamma$.

Let us now describe several important non-uniform stationary profiles
of the electric field. Consider in the first place the curves formed by
small triangles in Fig.\ \ref{IVjm}. They correspond to non-uniform
solutions with $E_j = E_L$ (the first fixed point of the
discrete mapping on the first branch of $v(E)$) for $1\leq j
\leq k$ (the number $k$ is shown on top of several such
curves in Fig.\ \ref{IVjm}). For $j>k$, $E_j$ ``jumps''
to the third branch of  $f(E;\gamma,J)$ and it approaches
the third fixed point of the discrete mapping ($1\rightarrow 3$
jumps). See these iterations of the discrete mapping in Fig.\
\ref{f2}(c) for a typical value of $J$. As $\phi$ increases and
$k$ decreases, each branch in Fig.\ \ref{IVjm} corresponds to the
same steady field profile but with the domain wall
displaced one step backward (towards $x=0$), which means that a
new quantum well has lost the low field value. In the main body of
Fig.\ \ref{IVjm}, only one out of each three branches of stationary
profiles of this type are depicted. The inset shows all these
solution branches for $k$ between 17 and 25. Since they turn
out to be linearly stable (see the Appendix), multistability between
different stationary profiles corresponding to the same voltage
$\phi$ is possible. As $\gamma$ grows, the width of the $J$
interval for which these solutions exist increases, and so
does the length of the triangle curves in Fig.\ \ref{IVjm}.
Then coexistence of {\em more} than two stable stationary
filed profiles for the same $\phi$ is possible. This gives
rise to larger hysteresis cycles and memory effects.
{\em When do the stationary field profiles with
$1\rightarrow 3$ jumps exist?}\, As the sequence of Figs.\
\ref{f2}(a) to \ref{f2}(c) shows, the values of $\gamma$ and $J$
should allow the local minimum of $f(E;\gamma,J)$ to descend below the
first fixed point of the mapping. This is possible for
$\gamma >\gamma_1$ and $J\in (J_1(\gamma),\sqrt{\gamma}v(E_M))$, where
the minimum of $f(E;\gamma,J)$ coincides with the first
fixed point at the current $J_1 (\gamma)$.
The branches of stationary solutions with $1\rightarrow 3$
jumps end at $J = \sqrt{\gamma} v(E_M)$ when the
fixed points on the first and second branches of the discrete
mapping coalesce and disappear.

Consider now the curves with small squares (and their
long-dashed continuations) in Fig.\ \ref{IVjm}. They
correspond to stationary solutions with $E_j = E_L$, which is now
the second fixed point of the discrete mapping, for the first
$k$ quantum wells. For $j>k$, $E_j$ jumps to the third branch
of $f(E;\gamma,J)$ ($2\rightarrow 3$ jumps). The short-dashed
lines in Fig.\ \ref{f2}(c) represent the iterations of the
discrete mapping leading to one of these solutions. For a given
$\gamma$, {\em for which range of $J$ do the stationary field profiles
with $2\rightarrow 3$ jumps exist?}\, We shall argue (and have checked
 numerically) that the profiles with $2\rightarrow 3$ jumps exist for
$J\in (J_2(\gamma),\sqrt{\gamma}\, v(E_M))$, where $J_2(\gamma)$ is
determined from (\ref{jhomo}) and (\ref{zero}) below. From
Fig.\ref{f2}(b) we see that these profiles appear at a value of $J =
J_2(\gamma)$ where the slope of $f(E;\gamma,J)$ at the second fixed
point of the mapping is zero. We observe in Fig.\ \ref{IVjm} that all
$2\rightarrow 3$ solution branches taper down to a point\cite{point}
$\phi_2(\gamma)$ for $J = J_2(\gamma)$. We find $\phi_2(\gamma)$ and $J
= J_2(\gamma)$ by inserting (\ref{jhomo}) into the condition that the
slope of $f(E;\gamma,J)$ be zero:
\begin{equation}
v + 2\,\sqrt{\gamma}\, v' = 0 \, . \label{zero}
\end{equation}
This equation has one solution corresponding to the minimum of
$f(E;\gamma,J)$ which determines $\phi_2(\gamma)$. Substituting
$\phi_2(\gamma)$ into (\ref{jhomo}) we obtain $J_2(\gamma)$. The
smallest possible $\gamma$ for which (\ref{zero}) has solutions is
$\gamma = \gamma_1 \simeq 5.2 \times 10^{-3}$: Non-uniform stationary
profiles exist for $\gamma > \gamma_1$. As we indicate in the Appendix,
the curve (\ref{zero}) bounds the region
inside which the uniform stationary solution is linearly unstable.
As in the case of the stationary field profiles with
$1\rightarrow 3$ jumps, only one out of three branches corresponding
to profiles with $2\rightarrow 3$ jumps is depicted in Fig.\ref{IVjm}.
Clearly many more stationary solutions can be constructed with the
help of the discrete mapping, and we have depicted only the more
significant ones in Fig.\ \ref{IVjm}. We want to point
out that all the stationary branches are connected: they do
not disappear as it could be wrongly inferred from
our Fig.\ \ref{IVjm}. For example, in Fig.\ \ref{f2}(c) we have
shown (long-dashed line) the mapping corresponding to a
stationary field profile different from the pure $1\rightarrow 3$
(continuous lines) and $2\rightarrow 3$ (short-dashed lines) jumps.
For these (unstable) solutions, $E_{k+1}$ jumps to the {\em second}
branch of $f(E;\gamma,J)$ (instead of jumping directly to the third
branch) before continuing toward the third fixed point of the mapping.
Clearly the branches corresponding to these solutions coalesce
with the $1\rightarrow 3$ branches at $J = J_1(\gamma)$ and
explain the disappearance thereof.
\bigskip

\noindent {\em Remark}. We have found stable non-uniform stationary
states having two domains with electric fields $E_L$ and $E_H$
separated by a domain wall. These fields are:

\begin{itemize}
\item $E_L<E_M$ (curves formed by small triangles in Fig.\ref{IVjm})
or $E_L\in (E_M,E_m)$ (curves formed by small squares in
Fig.\ref{IVjm}).
\item $E_H\in (E_m,E_{M'})$.
\end{itemize}
Thus while the first domain may have an electric field below or above
the SRT value $E_M$, the field at the second domain is always below the
second SRT value $E_{M'}$. Notice that this happens independently of
the applied bias for $\phi\in (E_M,E_{M'})$. That the field at the
domains does not necessarily coincide with the field for which there is
resonant tunneling is in agreement with experimental observations,
\cite{merlin}, and previous theories, \cite{laik}. For applied bias on
$(E_M,E_{M'})$, the average of the stationary
current at the branches representing non-uniform states is independent
of the bias, which explains the ``flat'' shape in Fig.\ \ref{IVjm}.
The experimental data present both flat plateaus and rising profiles
in the I--V characteristic curves of the samples. \cite{merlin,grahn90}

\section{Laser power vs. applied bias ``phase diagram''}
\label{sec-phase}

With the aim of interpreting the experimental results, we
have to delimit the intervals of the parameters $\gamma$ and $\phi$
where stable solutions of the types described in the preceding Section
may be found. A discussion of their linear stability is postponed to
the Appendix, but this information together with that provided by the
discrete mapping of the the preceding Section allows us to draw the
$\gamma - \phi$ ``phase diagram'' of Fig.\ \ref{phase}. The continuous
line in Fig.\ \ref{phase} bounds the region in which stable non-uniform
solutions with domains exist. The flat bottom of this region is at
$\gamma = \gamma_1$, the photoexcitation below which non-uniform
stationary states do not exist. The continuous line in Fig.\
\ref{phase} has been obtained numerically by the following procedure:

{}From Fig.\ \ref{IVjm} we see that the branches of stable non-uniform
solutions that yield the minimum and maximum values of $\phi$
are the two extreme small triangles. We have used the discrete
mapping (\ref{13}) to find with precision those extreme values of $\phi$
as functions of $\gamma$. For a given $\gamma$, the range of $J$ for
which the discrete mapping (\ref{13}) allows for stable non-uniform
solutions of any type (with $1\rightarrow 3$ or $2\rightarrow 3$ jumps)
has been determined following the procedure explained in Section II.
Notice that the minimum value of $\phi$ (in the continuous line of the
phase diagram Fig.\ \ref{phase}) is determined by taking the
smallest $J = J_1(\gamma)$ for which it is possible to find a solution
of (\ref{13}) and (\ref{5}) having 39 wells with field $E_L$
at the lowest fixed point and 1 well with a higher electric field.
On the other side, the maximum value of $\phi$ on the continuous line
of Fig.\ \ref{phase} is reached at the end of the last small triangle
branch of Fig.\ \ref{IVjm}. This value of $\phi$ is found by letting
$E_0 = E_1 = E_L$ be the second
fixed point of (\ref{13}) at $J = \sqrt{\gamma}\, v(E_M)$\, (the value
of $J$ for which the two first fixed points of (\ref{13}) coincide) and
then using (\ref{13}) to find the other values $E_j > E_L$ with
$j > 1$.

		Inside the continuous line in Fig.\ \ref{phase} we have depicted with
a dashed line the curve (\ref{zero}). As we explain in the Appendix, the
uniform solution is linearly unstable inside this curve. Between both
curves in Fig.\ \ref{phase}, both the uniform solution and at least one
branch (often several ones) of non-uniform solutions are linearly
stable for a given value of $\phi$. As discussed in Section II, we can
easily visualize multistability with the help of Fig.\ \ref{IVjm}.

\section{Simulation}

{}From the experiments  we have information about the
{\em steady state} of the problem through the current--voltage
characteristics, and about the {\em dynamical} aspects thereof through
the time evolution of the PC and of the PL.
In our model we have direct access to the time evolution of the
PC, proportional to $J(t)$, while we would need to supplement
our model to be able to obtain the time evolution of the PL.
We shall consider the evolution of the $J(t)$ and of the electric field
and charge profiles in this Section, while we will give qualitative
arguments about the evolution of the PL in the next Section.

In the previous stationary study we have found a very rich $\gamma-J$
phase diagram, showing multistability and hysteresis. We have seen
that our model accounts in a very satisfactory way for the static
properties observed experimentally. The picture will be substantially
completed by describing the time evolution of solutions of our model,
which will explain most of the dynamical
experimental results associated with the PC \cite{merlin}.
To extract time-dependent magnitudes from our model we have solved
numerically  the system of first order nonlinear equations, by
means of a standard fourth order Runge-Kutta method. For a fixed
value of voltage and laser power  we turn on
the laser at $t=0$. The initial condition for the fields and the charge
carriers are equations (\ref{11} - \ref{12}), except that we increase
the charge at the first quantum well to
mimic the effect of the laser, and then let the system evolve
according to Equations (\ref{1} - \ref{5}).

The analysis of the steady states in Section \ref{sec-steady} reveals
that at low laser power $\gamma < \gamma_1$\, ($\simeq 5.2\times
10^{-3}$ in our rescaled parameters), there is only one steady
solution: the uniform one, (\ref{11}--\ref{12}), which is linearly
stable (to be more precise, the same stands for the whole region
outside the full line of
the $\gamma - \phi$ phase diagram shown in Fig.\ \ref{phase}).
This result confirms the
usual assumption that the PC at low power laser is just a
rescaled plot of the velocity of the electrons versus the electric
field \cite{footnote}, and supports our choice of boundary conditions.
Above this threshold there is always a voltage interval (inside the
full line in Fig.\ \ref{phase}) where multiple steady states exist.
These states have step-like electric field profiles
as described above and some of them are linearly stable (see Fig.\
\ref{IVjm}). These solutions may even coexist with the uniform
one, which is unstable only inside the dashed line in Fig.\ \ref{phase}.

Nevertheless, from a dynamical point of view, the simulations reveal
an interesting behavior in the laser power range $0<\gamma<\gamma_2$
($\simeq 8 \times 10^{-3}$) for $\phi$ on a subinterval of
$(E_M,E_{M'})$ (for $\gamma = 0.016$, $1\leq \phi\leq 1.4$; see Fig.\
 \ref{phase}). The situation may be described as follows:

\subsection{$0<\gamma<\gamma_1$}
There is only
one steady state: that corresponding to an uniform profile of the
electric field. The linear stability analysis reveals that this steady
state is stable. However, if the disturbance is large enough, the
numerical simulation shows that, after a short time, a non-steady
step-like electric field profile is formed. The domain wall of this
profile oscillates back and forth in time about a fixed value of the
position. As it oscillates, the domain wall changes its width, so that
it spreads out over several wells, which makes easier the possibility of
experimental detection. The electric field at the two domains in the
profile also
oscillates in time for a few periods before the oscillations stop (see
Fig.\ \ref{movie_1} to visualize this process). Then there follows a
very slow modification of the step-like profile which ends up in the
uniform steady state. We call the step-like electric field
profile that remains after the PC oscillations cease, the
{\em metastable} state or the metastable domain formation. In Fig.\
\ref{movie_2}, we have plotted the PC versus time for
moderate and long times (inset), showing the oscillations and the
final decay of the metastable state. Notice that the oscillations are
appreciable only for $\gamma\geq 2 \times 10^{-4}$. Thus for
$0 < \gamma < 2 \times 10^{-4}$ there is {\em overdamped} formation of
a step-like field profile followed by a slow relaxation towards the
uniform steady state.

\subsection{$\gamma_1<\gamma<\gamma_2$}
There are true non-uniform steady states some of which are stable.
The numerical simulations show a pattern
analogous to that described above without the slow drift towards the
uniform state: after the PC oscillations cease, a non-uniform
steady state with two domains of the electric field is reached.

\subsection{$\gamma>\gamma_2$}
The damping is so strong
that there are no oscillations of the PC: the step-like
electric field profile is reached in a monotone fashion.
\bigskip

These results are visualized in Fig.\ \ref{7.5} in which the evolution
of the current with time is represented for increasing values of
$\gamma$ in a range that covers the three situations just described.
The region of the phase diagram (Fig.\ \ref{phase}) where oscillations
-- about either stable or metastable non-uniform solutions -- can be
observed has been represented
by a shaded rectangle: $2 \times 10^{-4} <\gamma < 8\times 10^{-3}$ and
$1.0<\phi<1.4$, approximately. Outside this region, the damping about
uniform or non-uniform (meta)stable steady states is so strong that no
current oscillations may be observed. In Fig.\ \ref{figcharge} we have
plotted the
profiles of the electric field and of the electron and hole densities
corresponding to a maximum and a minimum of the PC during one
oscillation. Notice that the electronic charge oscillates from one side
of the domain wall to the other during a period of the PC
oscillation \cite{merlin}. These results provide a clear picture of
domain formation and posterior time evolution accounting for the
time-dependent oscillations of the PC.

\setcounter{equation}{0}
\section{An interpretation of the PL measurements}

The PC discussed in the previous section gives an integrated information
about our sample, and is by no means a clear indication of domain
formation. The experimental evidence of  domain formation is the
evolution in time of the PL spectrum which measures the light emitted
by the sample in the electron hole recombination processes. The time
resolved PL is a direct way to measure the time evolution of the
carrier distribution. A correct description of the PL spectra would
require explicit consideration of recombination light emission,
scattering and tunneling times in the SRT process and of experimental
resolution, which is out of the scope of the present work. However, we
would like to supplement our model with a few assumptions so as to get
a qualitative interpretation of the PL measurements. To do this, we
need to understand the processes involved on the basis of the
experimental information \cite{merlin}, which can be summarized as
follows:

\begin{enumerate}

\item For large enough laser power and applied voltage,
two PL peaks appear when the laser is switched
on, confirming the formation of electric field domains.
Due to the Quantum Confined  Stark Effect (QCSE) \cite{QCSE} we
can affirm that the peak that appears at lower energy corresponds
to higher value of the electric field than that at higher
energy. The intensity of each peak is proportional to the number of
electron-hole recombinations.

\item  As the applied voltage increases the intensity of the high
field peak (low energy) becomes greater but the position
of the peaks remains approximately the same.

\item The intensity of the peaks oscillates with time at the same
frequency as the PC oscillations.

\item  The intensity of the low field peak (high energy)
is greater for a maximum of the PC than for a minimum
and the reverse occurs for the high field.

\end{enumerate}

The two first points indicate that the electric field value at each
domain does not vary as the applied voltage increases: the domain wall
just moves toward the $p$ contact thereby increasing (decreasing) the
size of the high (low) field domain.
The third point shows that the electrons responsible for the
PL are the same ones that carry the current across the
sample.

All these points are highlighted in the following very simple formula
which relates the PL spectrum, $PL(\hbar \omega)$, to the carrier
densities and the electric field at each well obtained from
our model:

\begin{equation}
PL(\hbar \omega) = \tilde{r} \sum_{j=1}^N \{\tilde{n}_j\,\tilde{p}_j
\,\Delta(\hbar\omega-\Omega(\tilde{E}_j)) \} .  \label{PLformula}
\end{equation}

\noindent Here $\tilde{n}_j$ ($\tilde{p}_j$) is the electron (hole)
concentration at the well $j$, $\tilde{r}$ is the recombination
coefficient assumed to be field independent, and $\Omega(\tilde{E}_j)$
is the energy of the photon emitted in the recombination process at
the $j$-th well. Due to the QCSE, this energy depends on $\tilde{E}_j$,
and it can be easily calculated with a one-well model
\cite{merlin_priv}. The result shows an almost linear behavior of
$\Omega(\tilde{E})$ with slope $2 e \tilde{l}_w$ ($e$ is the electron
charge and $\tilde{l}_w$ is the width of the well). In
(\ref{PLformula}), $\Delta(x)$ is just a Lorentzian function that
counts the contribution of $\Omega(\tilde{E}_j)$ to the PL spectrum
at energy $\hbar\omega$ with a finite resolution. In writing the
expression (\ref{PLformula}) we have implicitly assumed that each well
is at local equilibrium during the recombination process as explained
in the Introduction.

{}From Fig.\ \ref{movie_2} we see that a maximum of the PC occurs
when the domain wall is almost vertical and the electric field at
most of the wells takes on either the low ($E_L\simeq E_M$) or the high
($E_H$) field values, for which the tunneling probability is higher
(cf.\ the curve $v(E)$). During one oscillation, the low field domain
(which contributes to the high energy peak in the PL spectrum) is
larger at the maximum of the PC than at its minimum. According to
(\ref{PLformula}) and our previous discussion, this should imply an
increase in the low field peak (high energy in the PL spectrum) for the
maximum of the PC. When the PC takes on its minimum value within one
oscillation, the domain wall is not so sharply defined and several
quantum well have a value of the electric field for which the tunneling
probability is small.

The number of wells that are in the high field domain increases with the
applied voltage, and therefore, the weight of the low energy peak (HF)
increases with the voltage, reflecting the growth of the HF domain.
In Fig.\ \ref{lumi} we show the PL spectrum for a maximum and a
minimum of the PC for the same values as Fig.\ \ref{movie_1} and
\ref{figcharge}. The result resembles the experimental spectrum but
it is clear that in order to get quantitative agreement a more
elaborate model of the sequential resonant tunneling processes and
electron--hole recombination is needed. This is out of the scope of
the present work.

\setcounter{equation}{0}
\section{Discussion}

We have analyzed a very simple model that addresses in a satisfactory
way the transport aspects of the problem which are dominant at the time
scales experimentally relevant (1 -- 100 ns \cite{merlin}): sequential
resonant tunneling between weakly coupled quantum wells,
photogeneration and recombination of electrons and holes, and
self-consistency through the Poisson equation. An important aspect of
our model is the {\em discrete} nature of its governing equations
(\ref{1}--\ref{4}). Recall that this crucial feature is imposed by the
separation between the relevant time scales of the above said transport
processes and those of intra- and intersubband scattering (cf. Section
I). Our results exhibit a wide variety of phenomena in agreement with
the experiments: multistability of the stationary solutions with its
consequences of hysteresis and memory effects \cite{grahn90},
time-dependent oscillations of the PC, domain and domain wall formation
\cite{merlin}. The metastability of the non-uniform stationary states
predicted from our model is compatible with current experimental data,
although longer observation times seem necessary for non-ambiguous
confirmation \cite{merlin}. Notice that the non-uniqueness of the
stationary solutions is a direct consequence of our equations and it
can be easily explained with the help of the multivalued discrete
mapping of Section \ref{sec-steady}. The non-uniform stationary field
profiles exist on the interval of applied bias $(E_M,E_{M'})$ which is
significantly {\em larger} than the NDR region $(E_M,E_{m})$. The
region of multistability in the current-voltage characteristic curve is
``flat'' (the average current does not change with the applied bias,
Fig.\ref{IVjm}). From our results it is clear that similar
multistability due to coexistence between domains I (at zero field) and
II (at field slightly below $E_M$) \cite{grahn90} may be obtained by
adding a local maximum at zero field to the $v(E)$ curve. A simple
modification of our Poisson equation and the equation for the holes may
allow the study of similar phenomena in doped superlattices, which will
be considered elsewhere \cite{dopedSL}.  It is important to notice that
our results hold for a generic curve $v(E)$ of the form discussed in
the paper and that we could fine-tune experimental data by changing the
quantitative features of $v(E)$ and the parameter values. For instance,
we could increase the size of the region where PC oscillations appear
(shaded rectangle in Fig.\ \ref{phase}) relative to $(E_M,E_{M'})$ by
displacing the local  minimum $E_m$ towards $E_{M'}$. As there would be
PC oscillations about non-uniform steady states with larger biases, the
domain wall during a period of these oscillations would be closer to
$x=0$ than in Fig.\ \ref{figcharge}. This would mean that we could
obtain a larger domain III. In turn, this would cause a larger HF peak
to appear in the PL spectrum of Fig.\ \ref{lumi} thereby improving the
agreement with experimental results.\cite{merlin} Similarly the
frequency of the PC time-dependent oscillations is a increasing
function of $\beta$: We could estimate the ratio between the tunneling
and recombination times by fitting the numerically obtained values of
the frequencies to the experiments.

An important question, raised in the Introduction, is the dependance
of our results with the boundary condition (\ref{bc}). Under this
condition, the uniform stationary solution is stable at low laser
power, which allowed us to identify the velocity curve with the I--V
characteristic curve (except for unimportant constant scaling factors).
At higher photoexcitation, the uniform stationary solution could become
unstable in favor of non-uniform two-domain solutions which agrees with
the usual experimental interpretation \cite{grahn90}. All these facts
are of course indirect evidence in favor of our choice (\ref{bc}).
Still one wonders whether other choices also yield results in
agreement with the known facts. Without entering in detailed modeling
of the contact and doped regions before the superlattice, one can
consider other reasonable choices: it is plausible that modeling the
region before the superlattice one obtains an increasing
current-voltage curve (at least for not too large bias). Since $E_0$ is
the average field before the superlattice, one could derive a formula
$E_0 = g(J)$ from that current-voltage curve and use it as an
alternative boundary condition instead of (\ref{bc}). Under this new
condition we can repeat our analysis of the stationary solutions:
Typically  $E_0 = g(J)$ would now be different from a fixed point of
the discrete mapping (\ref{13}), so that the uniform field profile
would not be a stationary solution of the discrete equations. The
closest to the uniform solution would be a monotone field profile where
a fixed point of  (\ref{13}) is reached after a few iterations. For
very long superlattices the difference with the results obtained with
(\ref{bc}) is small. Non-uniform two-domain solutions would still be
possible: after a few iterates $E_j$ would be close to the first fixed
point of the mapping and the rest of our analysis in Section
\ref{sec-steady} can be used to construct the non-uniform solutions. The
differences would of course be more noticeable for short superlattices.
If (for example) $E_0 = g(J)$ turns out to be small (think for instance
of a linear law $g(J) = r J$ with a small resistivity $r$), there would
be a region of the superlattice with a field smaller than that
corresponding to the low field domain of Section \ref{sec-steady}.
These differences should be experimentally testable, and in the absence
of such tests and given the present good agreement with known facts
\cite{merlin,grahn90}, we prefer to stick to the simpler situation
provided by the boundary condition (\ref{bc}). Would future data force
another choice of boundary condition upon us, our methodology could
obviously be used to analyze the corresponding problem.

\bigskip

Our theory compares favorably with previous ones:

Grahn et al \cite{grahn90}
considered a continuous Poisson equation that ignored the holes and
a stationary Amp\`{e}re's law with a drift velocity
consisting of a sum of delta functions. They imposed a step-like
stationary electric field profile as a solution and obtained
several qualitative features of the experimentally observed
domains. Their results share with ours the fact that the average of
the stationary current density at the branches representing
non-uniform states is flat (independent of the applied bias, Section
II). Dynamics and stability properties of non-uniform profiles were
not considered in Ref. \onlinecite{grahn90}.

Laikhtman's theory \cite{laik}
consists of a discrete Amp\`{e}re's law plus the condition that
the total voltage bias is constant. He did not consider the
carrier densities and equations for them (Poisson's law and a
rate equation for the holes). Thus the only coupling between
quantum wells in his theory comes from the bias condition
and the electric field in the wells may alternate between
values at the different branches of $v(E)$ without restrictions.
Then quite complicated non-monotone stable stationary field profiles
with several domains are possible, as indicated in Ref.
\onlinecite{laik}. Within Laikhtman's theory
these profiles cannot be eliminated: the only possible
correlation between domains would be caused by assuming coherent
tunneling transport between wells. Our model
yields monotone non-uniform stationary profiles in a natural way
without appealing to coherent tunneling. The number of possible
non-uniform stationary states is also much smaller than in Ref.
\onlinecite{laik}.

Korotkov et al's theory \cite{korotkov} corresponds to
quite different superlattices (slim) where the tunneling processes
are single-electron tunneling events and the charge is quantized.
Interesting effects are predicted in this limit: coexistence of
Bloch oscillations and oscillations due to single-electron
tunneling \cite{likh}.

Other theories are based upon drift-diffusion models and are unable to
reproduce important qualitative features present in the experiments
such as the multiplicity of steady states in the NDR region, and the
eventual damping of the oscillations with time \cite{leperson}. In
fact, our model resembles known drift-diffusion models in the continuum
limit (such as the one in Ref. \onlinecite{swt}) for which the
uniqueness of the steady state can be proved easily, at least in the
limit of small diffusion coefficient \cite{smsig} (these continuum
drift-diffusion models are also known to have Gunn effect oscillations
\cite{shawold,HB} among their possible solutions \cite{inma}, which is
not the case for Eqs.\ (\ref{1}--\ref{5})). Thus it seems that the
discreteness of our model is a crucial ingredient in reproducing both
important static and dynamical features of transport in a superlattice
made out of weakly coupled quantum wells. Work on derivation of
discrete drift models from microscopic ones is now in progress.

\acknowledgements

We thank Prof. R. Merlin and Dr. S.-H. Kwok for proposing the problem
of superlattice dynamics to us, and for many helpful discussions,
critical comments and collaboration. We thank Dr. O. M. Bulashenko,
Dr. G. Platero and Prof. S. W. Teitsworth for fruitful discussions and
collaboration on related topics and Dr. C. Jagels for helpful
discussions on the linear stability analysis. This work has been
supported by the DGICYT grant PB92--0248, by  the NATO traveling grant
CRG--900284, and by the EC Human Capital and Mobility Programme
contract ERBCHRXCT930413. One of us (J.A.C.) also acknowledges
financial support of the DGICYT grant PB91-0378.

\appendix

\section*{Linear stability of the steady solutions}

Linearization of Eqs. (\ref{1}--\ref{5}) about any steady solution
(denoted by $E^0_j$, $n^0_j$, $p^0_j$ and $J$) yields the following
system for the disturbances ($\hat{e}_j$, $\hat{n}_j$, $\hat{p}_j$ and
$\hat{J}$) from it:
\begin{eqnarray}
\hat{e}_j-\hat{e}_{j-1} & = & \hat{n}_j-\hat{p}_j \, ,\label{lin1} \\
\beta\,\frac{d\hat{e}_j}{dt} & = & \hat{J}-v(E^0_j)\,\hat{n}_j
-n^0_jv'(E^0_j)\,\hat{e}_j\, ,  \label{lin2} \\
\frac{d\hat{p}_j}{dt} & = & -n^0_j\hat{p}_j - p^0_j\hat{n}_j \, ,
\label{lin3}  \\
\sum_{j=1}^N\hat{e}_j & = & 0  \, , \label{lin4}  \\
\hat{e}_0 & = & \hat{e}_1 \, .  \label{lin5}
\end{eqnarray}
where, as usual, $j$ runs from 1 to N. The system above can be reduced:
we can eliminate the $\hat{n}$'s in favor of the $\hat{p}$'s and the
$\hat{e}$'s by using (\ref{lin1}) and (\ref{lin5}). Moreover, summing
(\ref{lin2}) for all $j=1,\ldots,N$, and using (\ref{lin4}), we obtain
$\hat{J}$ in terms of the same variables. Finally a system of $2N$
equations (\ref{lin2}--\ref{lin3}) for the $2N$ unknowns ($\hat{e}_j$,
$\hat{p}_j$, $j=1,\dots,N$) results. Time evolution preserves a
constant value for the sum of all the fields, but the condition that
this value be zero is to be imposed as an additional constraint to the
resulting system of equations. They can be written in matrix form as
$dx/dt = M\,x$,\, where we have defined the vector
$x=(\hat{e}_1,\dots,\hat{e}_N,\hat{p}_1,\dots,\hat{p}_N)^{\rm T}$
($\phantom{A}^T$ denotes the transpose). The eigenvalues of $M$ should
in principle determine the linear stability of the steady solutions; in
practice, however, one of the eigenvalues of $M$ is spurious and we
should eliminate it by taking into consideration the constraint
(\ref{lin4}).

We can find the exact eigenvalues for the uniform solution
(\ref{11}--\ref{12}). In this case, the matrix $M$ is
\begin{equation}
M=\left(\begin{array}{cc}  A_{ee} & A_{ep} \\ A_{pe} & A_{pp}
\end{array}\right)\, ,
\end{equation}
where
\begin{eqnarray}
A_{ee} & = & -\frac{v}{\beta N}A-\frac{v}{\beta}B-
\sqrt{\gamma}\frac{v'}{\beta}I + \frac{\sqrt{\gamma}\, v'}{\beta\,
N}F\, , \label{matAee}    \\
A_{ep} & = & \frac{v}{\beta N}F-\frac{v}{\beta}I   \, ,
\label{matAep}    \\
A_{pe} & = & -\sqrt{\gamma}B  \, , \label{matApe}    \\
A_{pp} & = & -2\sqrt{\gamma}I \, . \label{matApp}
\end{eqnarray}
Here $v$ and $v'$ denote $v(\phi)$ and $v'(\phi)$ respectively, $I$
is the $N\times N$ identity
matrix and $A$, $B$ and $F$ are the following $N\times N$ matrices:
\begin{equation}
A  =  \left(\begin{array}{ccccc}
  1    &    0   & \cdots &    0   &   -1   \\
\vdots & \vdots &        & \vdots & \vdots \\
\vdots & \vdots &        & \vdots & \vdots \\
\vdots & \vdots &        & \vdots & \vdots \\
  1    &    0   & \cdots &    0   &   -1
\end{array}\right)\, ,   \label{matA}
\end{equation}
\begin{equation}
B  =  \left(\begin{array}{ccccc}
  0    & \cdots & \cdots & \cdots &    0   \\
 -1    &    1   &    0   & \cdots &    0   \\
  0    &   -1   &    1   & \ddots & \vdots \\
\vdots & \ddots & \ddots & \ddots &    0   \\
  0    & \cdots &    0   &   -1   &    1
\end{array}\right)\, ,   \label{matB}
\end{equation}
\begin{equation}
F  =  \left(\begin{array}{ccc}
  1    & \cdots &   1    \\
\vdots &        & \vdots \\
  1    & \cdots &   1
\end{array}\right)\, .   \label{matF}
\end{equation}
The matrix $M$ has only four distinct eigenvalues, given by the
expressions:
\begin{eqnarray}
\lambda_0 & = & 0 \, , \label{lambda0} \\
\lambda_1 & = & -2\sqrt{\gamma} \, , \label{lambda1} \\
\lambda_{\pm} & = & -\sqrt{\gamma}\left(1+\frac{v'}{2\beta}\right)
-\frac{v}{2\beta}   \nonumber  \\
 & \pm & \sqrt{ \frac{v}{4\beta^2}(v+2v'\sqrt{\gamma})
+\gamma\left(1-\frac{v'}{2\beta}\right)^2 } \, . \label{lambdapm}
\end{eqnarray}
The two first eigenvalues are simple:
$\lambda_0$ is the spurious eigenvalue, corresponding to the
eigenvector $(1,\dots,1,0,\dots,0)^{\rm T}$ (notice that
$\sum_{j=1}^N\hat{e}_j\ne 0$ for this eigenvector) and therefore it
has to be ignored; $\lambda_1$ corresponds to the eigenvector
$(0,\dots,0,1,\dots,1)^{\rm T}$ and it is always real and negative.
The remaining two eigenvalues ($\lambda_+$ and $\lambda_-$) have
algebraic multiplicity $N-1$ each. It is not difficult to check that
they are always real and that at least one is always negative.
Accordingly, the stability of the uniform solution depends on the
sign of the other eigenvalue. It is simpler to consider the sign of
the product $\lambda_+\lambda_-$. The solution will be unstable
whenever this product is negative and neutrally stable when it is
zero, i.e. when
\[
v+2\sqrt{\gamma}v'=0\, .
\]

This equation is exactly (\ref{zero})!
Its
minimum  is $\gamma_1$, the photoexcitation below
which the uniform solution is the only existing stationary state.

It is clear that we could in principle ascertain the linear stability
of any steady solution of (\ref{1}--\ref{5}) by the method so far
explained. We would use any of the well known diagonalization
numerical routines on the matrix $M$ corresponding to the steady
solution. However these routines quite disastrously fail to determine
the exact value of the eigenvalue of $M$ because of the high
multiplicity thereof: the errors of the numerically
determined eigenvalues can be as high as 10\%, and, what is even
worse, they become artificially simple and sometimes even acquire a
spurious imaginary part. The fact that the matrix $M$ for the
uniform steady solution is not diagonalizable increases the
numerical difficulty. It is important to notice that the linear
stability of the uniform stationary solution is not affected by these
numerical errors: it depends only on the {\em sign} of the largest
eigenvalue and we have checked that this sign is given correctly by the
numerical code (we have always compared the results provided by the
code with the formula (\ref{lambdapm}) with the $+$ sign that yields the
largest eigenvalue exactly).

It could seem that these numerical troubles due to eigenvalue
degeneracy may not appear for domain-type steady solutions (which are
non-uniform). However, we do not have now {\em exact} means of checking
this conjecture at our disposal. Thus our numerical determination of
the linear stability of domain-type steady solutions has always been
confronted with the results of direct simulation of the full system of
equations whenever the comparison was possible. We have found that (as
in the case of the uniform solution) the linear stability
of the non-uniform steady states is correctly predicted by the numerical
determination of the eigenvalues. There is a	reasonable
agreement between the two numerical procedures for the real part of the
largest eigenvalue while there are large errors for the imaginary
part. Thus while the eigenvalue calculation gives acceptable
estimations of the {\em damping} of the photocurrent oscillations,
it provides poor estimations of the {\em frequency}. Hence
the stability properties of a given stationary state are well
determined by the numerical procedure and we have used it,
together with the simulations, to delimit the region of the
$\gamma-\phi$ phase diagram (Fig.\ \ref{phase}) inside which stable
non-uniform stationary solutions exist. Work on devising reliable
numerical codes to determine the oscillation frequency from the
eigenvalues is now in progress.

\begin{figure}
\caption[]{Different regions and processes that contribute to the PL
spectrum (see text). Only the conduction band is represented.}
\label{jumps}
\end{figure}

\begin{figure}
\caption[]{Velocity of the electrons as a function of the electric
field. It is modeled by two lorentzians centered at $E_M$ and
$E_{M'}$ which correspond to the sequential resonant tunneling
current between the quantum wells $j$ and $j+1$. At a field $E_M$,
there is SRT between the first level of the $j$-th well and the second
level of the $(j+1)$-th well, whereas at a field $E_{M'}$ there is SRT
between the second level of the $j$-th well and the third level of the
$(j+1)$-th well. (In different samples, the latter SRT process may
relate the first level of the $j$-th well and the third level of the
$(j+1)-$th well).}
\label{ve}
\end{figure}

\begin{figure}
\caption[]{Static characteristic curve: current ($J$) versus applied
bias ($\phi$) for laser power $\gamma = 0.016$ ($\beta$, which is only
relevant for the stability of the different solutions -- see
(\ref{lin2}) in the Appendix, is assumed to be 0.05). The solid line
corresponds to the stable uniform solution. In the NDR part of the
curve there is a region where
this solution becomes unstable (short-dashed line). The remaining
curves correspond to two-domain solutions. There are 39 branches (one
less than quantum wells), though only one every three branches has been
plotted for the sake of clarity. The inset enlarges a small region of
the figure showing the branches in full detail. Each branch has a
different number of wells (from 1 to 39) in the low field domain (the
number written on top of some of them). Every branch consists of two
curves, according to the way they are built with the discrete mapping
(\ref{13}) (see text for details): one plotted with triangles,
corresponding to jumps from the first to the third fixed points of
(\ref{13}), and another plotted with a combination of squares (stable
part) and long dashed lines (unstable part), corresponding to jumps from
the second to the third fixed points of (\ref{13}). Important features
to stress from this figure are: first, the coexistence, for a given
$\phi$, not only of several domain solutions but also of domain
solutions and the uniform solution (see also Fig.\ \ref{phase}),
something relevant to the appearance of hysteresis and memory effects,
and second, the average ``flatness'' of $J$ in the bias region between
the two local maxima of $v(E)$.}
\label{IVjm}
\end{figure}

\begin{figure}
\caption[]{The discrete mapping: $f(E;\gamma,J)$ and $E$ versus $E$ for
$\gamma = 0.016$ and five different values of $J$. {\em (a)} For small
$J$ (below the minimum of $\sqrt{\gamma}\, v(E)$), only one fixed point
of the discrete mapping appears; thus only uniform solutions are
allowed. {\em (b)} There are three fixed points of the discrete
mapping, but only $2\rightarrow 3$ jumps are allowed. {\em (c)} Now
both $1\rightarrow 3$ and $2\rightarrow 3$ jumps are allowed. The
continuous lines show the discrete mapping process used to construct
a ($1\rightarrow 3$) non-uniform stationary solution with two domains:
After remaining at the low field domain value $E_L$, the field jumps at
the $(m+1)$-th well to a different solution of $f(E;\gamma,J) = E_L$.
In successive iterations the electric field tends to its high-field
value $E_H$. The short-dashed lines show the same construction for a
$2\rightarrow 3$ solution, whereas the long-dashed lines correspond to
a different (unstable) type of $2\rightarrow 3$ solution explained in
the text. {\em (d)} Same as {\em (c)}, but showing the coalescence of
solutions with $1\rightarrow 3$ and $2\rightarrow 3$ jumps for $J =
\sqrt{\gamma}\, v(E_M)$. {\em (e)} For large $J$, above the first
positive maximum of $\sqrt{\gamma}\, v(E)$, there is only one fixed
point of the discrete mapping and the situation is the same as in
{\em (a)}: the only possible stationary solutions are uniform.}
\label{f2}
\end{figure}

\begin{figure}
\caption[]{Phase diagram of the model:
Laser power ($\gamma$) vs. voltage bias ($\phi$).
Stable non-uniform solutions with domains exist only inside the
continuous curve.
Inside the dashed line the uniform solution is unstable.
The shaded  rectangle approximately
delimits the region where oscillations of the PC (i.e., of
the domain wall of the electric field profile) can be obtained.}
\label{phase}
\end{figure}

\begin{figure}
\caption[]{Evolution of the domain wall in time for $\gamma=2\times
10^{-3}$ and $\phi=1.2$. {\em (a)} is a contour plot with time flowing
in the $y$-axis and the quantum well number in the $x$-axis. The
electric field contours go from 0.5 to 1.7 in 0.1 steps. At $t=0$
we start with the uniform solution. The perturbation is clearly
visible at the first well at short times. The creation and the
displacement of the domain wall can be visualized. At $t = 0.5$ the two
domains are already defined (the darker the figure the higher the
electric field) and the domain wall begins to oscillate. Most of the
wells have an electric field centered around the first maximum of
$v(E)$ (low field domain). In {\em (b)} we represent the electric field
profile at short times (domain formation) and in {\em (c)} the domain
wall oscillations.}
\label{movie_1}
\end{figure}

\begin{figure}
\caption[]{The PC ($J$) evolution for the same values of the
parameters as Fig.\ \ref{movie_1}. The transient before the
oscillations corresponds to the formation of the low field domain.
The oscillations of the domain wall induce those of the PC.
The maxima are reached when the domain wall is sharper. In the inset
it can be observed the small damping of the PC, as the system
slowly approaches the stable uniform solution. This damping disappears
for large enough laser power (see text).}
\label{movie_2}
\end{figure}

\begin{figure}
\caption[]{Evolution of the PC, $J$, for different values of
the photogeneration, $\gamma$ (in units of $10^{-3}$), which increases
from bottom to top of the figure, as indicated at the right margin.
The voltage ($\phi=1.1$) is the same for all the curves.}
\label{7.5}
\end{figure}

\begin{figure}
\caption[]{(a) Electric field profile and (b) charge at each well
corresponding to a maximum (full line) and a minimum (dashed line) of
the PC.}
\label{figcharge}
\end{figure}

\begin{figure}
\caption[]{Oscillations of the PL
spectrum corresponding to Fig.\ \ref{movie_2}.}
\label{lumi}
\end{figure}


\begin{references}
\bibitem{point}
At this point all the $N-1$ non-uniform solution branches (with $2
\rightarrow 3$ jumps) coalesce with the uniform solution, which has
a zero eigenvalue of multiplicity $N-1$. For a given $\gamma >
\gamma_1$, the difference between the biases corresponding to two
contiguous branches of these non-uniform solutions in Fig.\ \ref{IVjm}
(with the same $J$) is equal to $(E_N - E_L)/N$, where the $E_j$
correspond to the solution branch with larger bias with $E_j = E_L$
for $1\leq j\leq k$. Let now $0 < (J - J_2) \ll 1$. Then it is
straightforward to check that the difference between the biases of
contiguous non-uniform solution branches is \,
$O\left( (J - J_2)^{2^{-(N-k-1)}}/N\right) $.
Clearly this difference shrinks to zero at $J = J_2$ for any fixed $N$.
\bibitem{footnote}
Notice that while we have chosen the shape of $v(E)$
to reproduce this fact, the uniqueness and stability of this solution at
low laser power is a result of our theory.
\bibitem{dopedSL}
For doped superlattices we do not need the photoexcitation to
generate carriers. We have found that the corresponding discrete
equations have persistent and faster time-dependent oscillations of
the current, confirming the damping role of the processes
involving the holes.
\end{references}
\end{document}